\documentclass{elsart}
\usepackage{amssymb}
\usepackage[german,english]{babel}
\usepackage[intlimits,sumlimits,namelimits]{amsmath}
\usepackage[dvips]{epsfig}
\usepackage[dvips]{color}
\usepackage{amssymb,ifthen,shadow}
\usepackage{fancyhdr}
\usepackage[dvips]{epsfig}

\usepackage[dvips]{epsfig}

\newcommand{\beq}{\begin{equation}}
\newcommand{\eeq}{\end{equation}}

\def\bar{\begin{array}}
\def\ear{\end{array}}

\def\newline{\hfil\break}
\def\newpage{\vfill\eject}

\def\NP{ Nucl. Phys.}

\def\PREV{ Phys. Rev.}
\def\PREP{ Phys. Rep.}

\def\PL {Phys. Lett.}
\begin{document}

\begin{frontmatter}

\title{Self-consistent treatment of the self-energy in nuclear matter}

\author{Kh.\ Gad\corauthref{khalaf}} {\bf and E.M.\ Darwish}
\corauth[khalaf]{e-mail: kha92@yahoo.com}

\address{Physics Department, Faculty of Science, South Valley University, 
  Sohag, Egypt}

\begin{abstract}
The influence of hole-hole propagation in addition to the conventional
particle-particle propagation, on the energy per nucleon and the momentum
distribution is investigated. The results are compared to the
Brueckner-Hartree-Fock (BHF) calculations with a continuous choice and 
conventional choice for the single-particle spectrum.
The Bethe-Goldstone equation has been solved
using realistic $NN$ interactions. Also, the structure of nucleon self-energy in
nuclear matter is evaluated. All the self-energies are calculated
self-consistently. Starting from the BHF
approximation without the usual angle-average approximation, the effects of
hole-hole contributions and a self-consistent treatment within the framework
of the Green function approach are investigated. Using the self-consistent
self-energy, the hole and particle
self-consistent spectral functions including the particle-particle and hole-hole ladder
contributions in nuclear matter are
calculated using  realistic $NN$ interactions. We found that, the difference
in binding energy
between both results, i.e. BHF and self-consistent Green function, is not large. 
This explains why is the BHF ignored the 2h1p contribution. 
\end{abstract}

\end{frontmatter}

\section{Introduction} 

It is one of the fundamental issue in nuclear physics to
evaluate the nuclear matter binding energy and saturation properties, starting
from a realistic
nucleon-nucleon ($NN$) interaction without any adjustment of a free parameter 
\cite{day,spr,hjo}. The solution of this
problem is of great interest, because the understanding of the binding energy of nuclei
is one of the basic problems of nuclear physics and physics in general. Because of the presence
of strongly repulsive components in the short-range part of the $NN$
interaction \cite{dewulf,fromel,bamonaco}, the $NN$ scattering correlation has a predominant
importance in the nuclear matter calculation. The solution of this problem is
also very important because the nuclear many-body problem is one of the most
challenging testing grounds for many-body theories of quantum systems. Last
not least, however, a microscopic understanding of the binding energy of normal
nuclei, which is based on a realistic model for the $NN$ interaction, shall also
provide a reliable prediction for the equation of state of nuclear matter at
densities beyond the saturation density of normal nuclei, which is required
for the study of astrophysical objects like the explosion of supernova and the
structure of neutron stars. 

Recently, Various approximation schemes have been developed to describe the correlations
which are induced into the many-nucleon wave function by the strong short-range and tensor
components of such a realistic $NN$ force. For a recent review on such methods
see, e.g., Refs. \cite {bald,mut}. The resulting correlation functions are the key
input for the study of the absorption of real and virtual photons by pairs of
nucleons as it is observed in ($\gamma$, $NN$) or (e, $e^{\prime}$ $NN$)
reactions. One of  the most popular approximation schemes, which is frequently
used in nuclear physics, is the Brueckner-Bethe-Goldstone (BBG)
theory \cite{day} which has been one of the tools for solving the many-body problem 
of nuclear matter for many years \cite{mut,brue,bethe,haftel}. The results of BHF calculations depend on the choice
of the single particle potential, $U(k)$. In the conventional choice, $U=0$
for $k>k_{F}$, and $U$ is the self-consistent BHF potential for $k<k_{F}$. The
alternative "continuous" choice has been proposed in \cite{jeukenne} for
which $U$ is again the self-consistent BHF potential, but it extends to
$k>k_{F}$. This continuous choice leads to an enhancement of correlation
effects in the medium and tends to predict larger binding energies for nuclear
matter than the conventional choice.

In this paper, special attention will be paid to study the self-consistent
self-energy. Note, that all the self-energies in this paper will be calculated
self-consistently. The self-energy of a nucleon contains all the information
necessary to obtain occupation probabilities, quasi-particle strength and
broadening features which alternatively can be visualized in terms of hole and
particle spectral functions.

It is the purpose of this paper to investigate whether an improven and more
consistent treatment of the many-body system, still based on nonrelativistic
dynamics and two-body forces, can provide new information on the saturation
problem of nuclear matter. A self-consistent treatment of the Green's
functions formalism leads to the so called self-consistent Green's function
(SCGF) theory \cite{ramos1,ramos2,ramos3,dew,dick,vond,gad1}. Attempts have been made to employ the
technique of a self-consistent evaluation of Green's functions
\cite{kada,krae} to the solution of the nuclear many-body problem. This method
offers various advantages: (i) The single-particle Green's function contains
detailed information about the spectral function, i.e. the distribution of
single-particle strength, to be observed in nucleon knock-out experiments, as
a function of missing energy and momentum. (ii) The method can be extended to
finite temperatures \cite{fromel,frmuther}, a feature which is of interest for the study of the nuclear
properties in astrophysical environments. (iii) The BHF approximation, the
approximation to the hole-line expansion 
which is commonly used, can be
considered as a specific approximation within this scheme. Attempts have been
made to start from the BHF approximation and include the effects of the
hole-hole scattering terms in a perturbative way \cite{koh1,koh2}. There 
is a good reason to believe that a
more complete treatment of the nucleon self-energy in the full framework of
Green-function theory will ultimately help resolve long standing questions
about nuclear saturation. Via this study we will be able to give an answer to
a fundamental question regarding reliability of the BHF theory.   
Application of SCGF theory has proven useful for
understanding the location and quantity of single-particle strength found in
the electron-scattering experiments. The focus of this paper is to apply SCGF
theory to nuclear matter to study the nucleon properties. 

After this introduction we will discuss some features of the self-consistent
BHF approach for the self-consistent self-energy in Sec. 2. The effect of the
hole-hole terms in the self-consistent self-energy and a self-consistent
Green-function will be presented in Sec. 3. The self-consistent spectral
function will be presented in Sec. 4. The main conclusions are
summarized in the final section.

\section{BHF approximation} 

One of the central equations to be solved in the BHF approximation is the Bethe-Goldstone
equation. It is given by
\begin{eqnarray}
<k^{\prime}_{1}k^{\prime}_{2}| G(\omega
)|k_{1}k_{2}>&=& <k^{\prime}_{1}k^{\prime}_{2}|v|k_{1}k_{2}>+ 
\sum_{k_{3}k_{4}}<k^{\prime}_{1}k^{\prime}_{2}|v|k_{3}k_{4}>
\nonumber\\ &&
\times
\frac{Q(k_{3},k_{4})}{\omega-E(k_{3},k_{4})}
<k_{3}k_{4}| G(\omega
)|k_{1}k_{2}>\mbox{,}\label{eq:betheg} 
\end{eqnarray}
where $v$ is the bare potential, $\omega$ denotes the starting energy, $Q$ is the Pauli operator which requires the
nucleon momenta to be outside the Fermi sea, E is the sum of the two
single-particle energies inside nuclear matter given by 
\begin{equation}
E(k_{1},k_{2})=\varepsilon_{k_{1}}+\varepsilon_{k_{2}}\,,
\end{equation} 
with the single particle energies 
\begin{equation}
\varepsilon_{k} = \frac{k^2}{2m} +Re \hspace{1mm}  \Sigma^{BHF} 
(\vec k,\omega=\varepsilon_{k})\,,\label{eq:bhf1}
\end{equation}
where the self-energy of a nucleon in nuclear matter with momentum $\vec k$ is
determined by the self-consistent equation
\begin{equation}
\Sigma^{BHF}(\vec k,\omega)=\int d^3k^{\prime}<\vec {k} \vec {k^{\prime}} |
G(\varepsilon_{k}+\varepsilon_{k^{\prime}})|\vec {k} \vec {k^{\prime}} > n_{0}(\vec k^{\prime})\,,\label{eq:selfbhf}
\end{equation}
with the occupation probability of a free Fermi gas with a Fermi momentum $k_F$
\begin{equation}
n_0 (\vec k^{\prime}) = \left\{ \begin{array}{ll} 1 & \mbox{for}\; |\vec k^{\prime}| \leq k_F \\
0 & \mbox{for}\; |\vec k^{\prime}| > k_F \end{array}\right.\label{eq:occ0} \,.
\end{equation}
In the BHF approximation, (\ref{eq:betheg}), (\ref{eq:bhf1}), and (\ref{eq:selfbhf}) are
solved self-consistently.

\begin{figure}
\begin{center} 
\epsfig{file=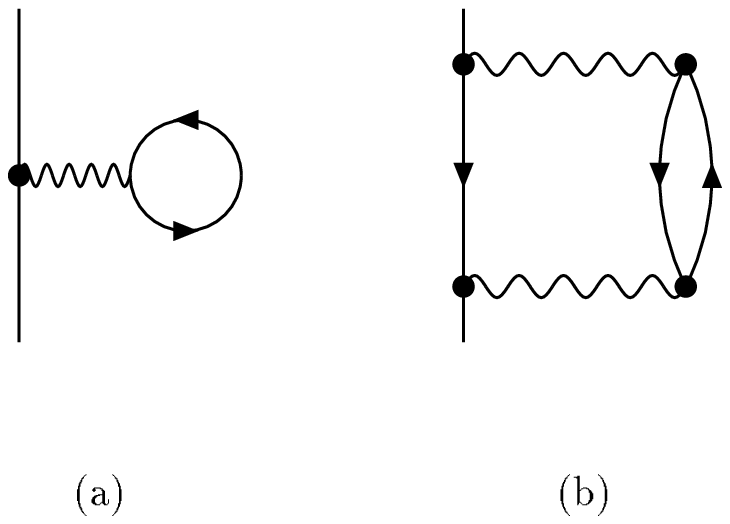,scale=1.,angle=0}
\end{center}
\caption{Diagrams contributing to the nucleon self-energy. The wavy
  lines indicate the Brueckner G-matrix. (a) The left diagram is the 
standard Brueckner approximation for the nucleon
self-energy (\protect{\ref{eq:selfbhf}}). 
(b) This diagram takes into account transitions to one-particle
two-hole states. It contributes to both real and imaginary parts of the
self-energy in the considered energy and momentum range. \label{1}}
\end{figure}

The definition of the single-particle potential or the self-energy can be
extended beyond the BHF approximation. In order to discuss such an extension we
want to consider various contributions, which are represented by the diagrams or
series of diagrams displayed in Fig. \ref{1}.  These
terms are summed up in a self-energy as can  be written schematically by
\begin{eqnarray}
\Sigma(k,\omega)& =& \Sigma^{HF}(k)+\Delta
\Sigma^{2p1h}(k,\omega)+\Delta \Sigma^{2h1p}(k,\omega)
\nonumber\\
& = &\Sigma^{BHF}(k,\omega)+ \Delta \Sigma^{2h1p}(k,\omega)\,,
\end{eqnarray}
where the contribution due to two-particle one-hole terms 
$\Delta \Sigma^{2p1h}$ is complex for $\omega > \varepsilon_{F}$
while the two-hole one-particle term $\Delta \Sigma^{2h1p}$ is complex for 
$\omega < \varepsilon_{F}$. The imaginary part of the BHF self-energy  vanishes for $\omega <
\varepsilon_{F}$. 

Real and imaginary parts of the self-energy are related to each other by a
dispersion relation of the form \cite{mahaux,wimo} \\
\begin{equation}
Re \Sigma^{BHF}(k,\omega) = \Sigma^{HF}(k) + \frac{1}{\pi}
\int_{-\infty}^\infty \frac{Im \hspace{1mm}
\Sigma^{BHF}(k,\omega')}{\omega'-\omega}d\omega'\,.\label{eq:disper1}
\end{equation}
\begin{equation}
Re \Delta\Sigma^{2h1p}(k,\omega) = \frac{1}{\pi}
\int_{-\infty}^\infty \frac{Im \hspace{1mm} \Delta
\Sigma^{2h1p}(k,\omega')}{\omega'-\omega}d\omega'\,,\label{eq:disper2}
\end{equation}

The self-energy of the Hartree-Fock term is real and independent of
$\omega$. It is given as \cite{ramosphd}
\begin{equation}
\Sigma^{HF}(k)=\sum_{k^{\prime}}<kk^{\prime} \mid V \mid kk^{\prime}> n
  (k^{\prime}) \mbox{,}
\end{equation}
where $n(k^{\prime})$ is the single particle occupation probability. The importance of the
Hartree-Fock term is that it represents the simplest solution to the many body problem.

\begin{figure}
\begin{center}
  \epsfig{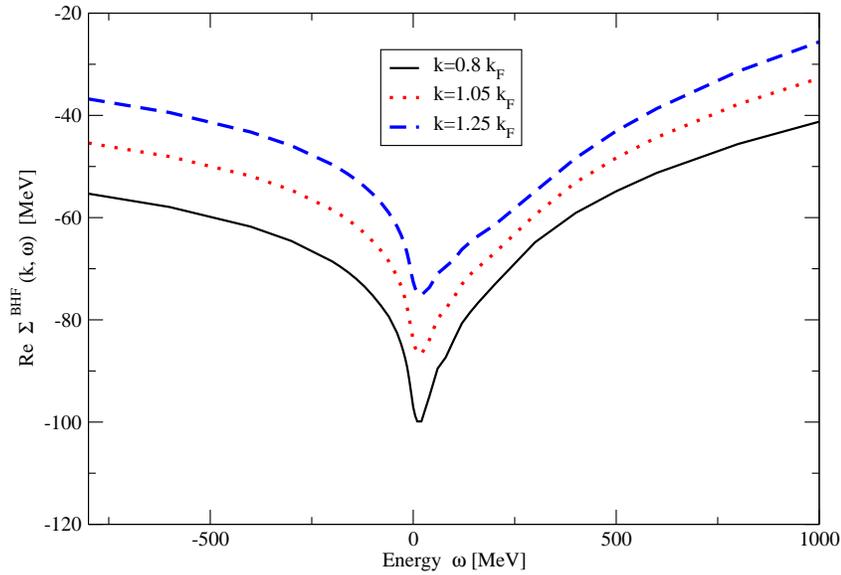}
\end{center}
\caption{The real part of the BHF self-consistent self-energy as a function of
  the energy $\omega$ (see (\protect{\ref{eq:selfbhf}})) for symmetric 
nuclear matter with Fermi momentum $k_F$ = 1.36 fm$^{-1}$ evaluated for the
CD-Bonn interaction, calculated for various momenta $k$. 
\label{2a}}
\end{figure}

\begin{figure}
\begin{center}
  \epsfig{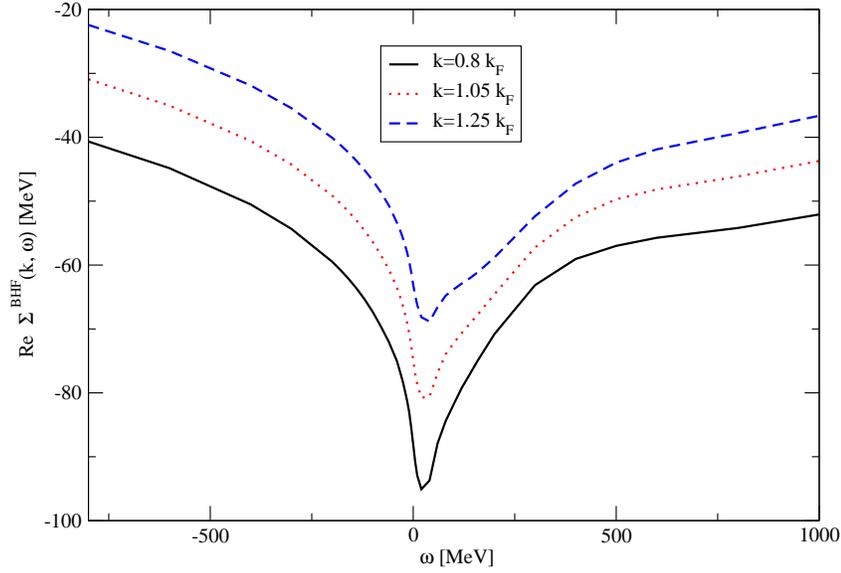}
\end{center}
\caption{The real part of the BHF self-consistent 
self-energy as a function of the $\omega$ (see (\protect{\ref{eq:selfbhf}})) for  
nuclear matter with Fermi momentum $k_F$ = 1.36 fm$^{-1}$ evaluated for the
Bonn C potential, for various momenta $k$. \label{2b}}
\end{figure}

The energy dependence of the real part of the BHF self-consistent self-energy 
(see (\ref{eq:selfbhf})) is visualized
in  Figs. \ref{2a} and \ref{2b}, for nuclear matter with Fermi momentum $k_{F}=1.36$
fm$^{-1}$. It is evaluated for the CD-Bonn \cite{mach} and Bonn C potential \cite{machleidt}, for various momenta
$k$. This real part displays a pronounced minimum at energies
around the Fermi energy. These results have been obtained after establishing
self-consistency for the single-particle spectrum according to (\ref{eq:bhf1}) 

\begin{figure}
\begin{center}
  \epsfig{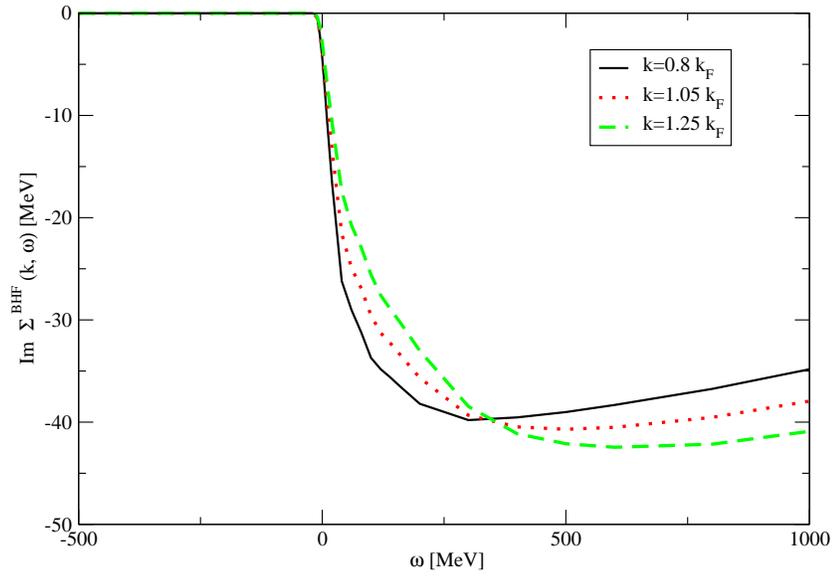}
\end{center}
\caption{The imaginary part of the BHF self-consistent self-energy as a
  function of $\omega$ (see (\protect{\ref{eq:selfbhf}})) for  
nuclear matter with Fermi momentum $k_F = 1.36$ fm $^{-1}$ calculated for the
CD-Bonn interaction, calculated for various momenta $k$. \label{3a}}
\end{figure}

\begin{figure}
\begin{center}
  \epsfig{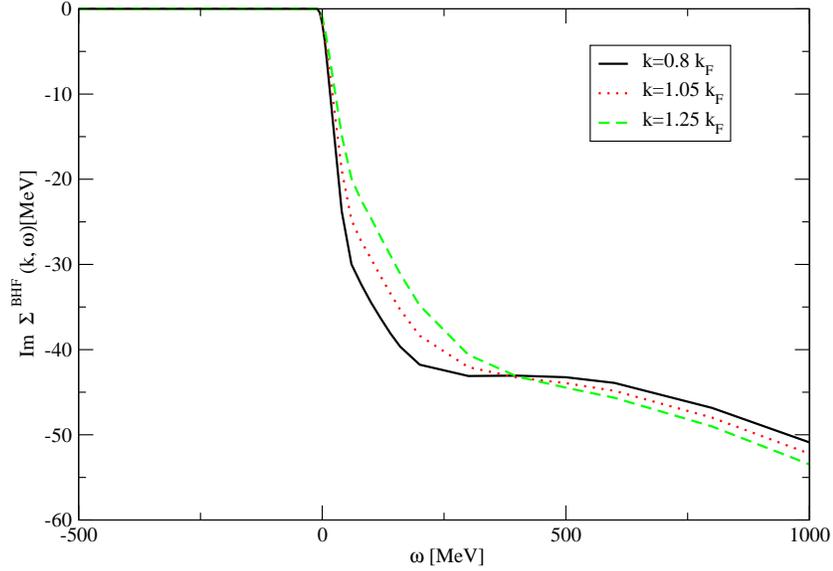}
\end{center}
\caption{The imaginary part of the BHF self-consistent self-energy for  
nuclear matter with Fermi momentum $k_F = 1.36$ fm $^{-1}$ calculated for the
Bonn C interaction, calculated for various momenta $k$. \label{3b}}
\end{figure}

Figs. \ref{3a} and \ref{3b} show the
imaginary part of the BHF self-consistent self-energy as a function of
$\omega$ evaluated for the CD-Bonn (Fig. \ref{3a}) and Bonn C (Fig. \ref{3b}) potentials, respectively, for different values of
momenta $k$. We see that the imaginary part is
identical zero for energies $\omega$ less than $\varepsilon_k - \varepsilon_F$,
as can be seen from (\ref{eq:bhf1}), and yield
non-negligible values up to very high energies.
Although not explicitly shown in Figs. \ref{3a} and \ref{3b}, Im$\Sigma^{BHF}(k,\omega)$
vanishes at high energies which is a consequence of the vanishing of the imaginary
$G$ matrix elements at very large values of the energy parameter. This result
is general for any soft-core interaction, although the softness of the core
directly influences the energy for which the matrix elements become negligible.
Considering the results for the imaginary part of the self-consistent self-energy, which are
displayed in  Figs. \ref{3a} and \ref{3b}, it is clear from the
dispersion relation (\ref{eq:disper1}) that the real part of the self-energy is identical to the
corresponding HF single-particle potential $\Sigma^{HF}(k)$ in the limit $\omega \to
-\infty$. This real part gets more attractive with increasing $\omega$ until
one reaches  values of $\omega$ at which the imaginary part is different from
zero. The self-energy turns less attractive at higher energies, which leads to
a pronounced minimum at energies $\omega$ slightly above the Fermi energy. 

The conventional choice has been to ignore self-energy contributions for the
particle states completely and approximate the energies by the kinetic energy
only. This conventional choice for the single-particle spectrum, however, is
not very appealing as it leads to a gab at the Fermi surface: the propagator
for single-particle states with momenta below the Fermi momentum $k_{F}$ is
described in terms of a bound single-particle energy while the corresponding
spectrum for the particle states starts at the kinetic energy for the momentum
$k_{F}$, so the potential in this case is given
by \cite{bbp}
\begin{equation}
U(k)=
  \begin{cases}
  \sum_{k^{\prime} \leq k_{F}} <k k^{\prime} \mid G(\omega= \varepsilon_{k}+\varepsilon_{k^{\prime}}) \mid k k^{\prime}>_{A} :\hspace{7mm}       k \leq k_{F}  \\
  0 \hspace{78mm}             k >k_{F}
  \end{cases}\,.\label{eq:conv}
  \end{equation}
  Alternatively, a continuous choice for the potential can be made \cite{machleidt}
  by defining
   \begin{equation}
  U(k)=Re \sum_{{k^{\prime}} \leq k_{F}} <k k^{\prime} \mid G(\omega= \varepsilon_{k}+\varepsilon_{k^{\prime}}) \mid
  k k^{\prime}>_{A} \mbox{,} \label{eq:cont}
  \end{equation}
 for all states $k$ below and above the Fermi surface. This leads to a
 spectrum that is continuous at the Fermi momentum. Where $\mid
 kk^{\prime}>_{A}\hspace{1mm}$ = $\hspace{1mm} \mid kk^{\prime}>$-$\mid k^{\prime}k>$.
 
\begin{figure}
\begin{center}
  \epsfig{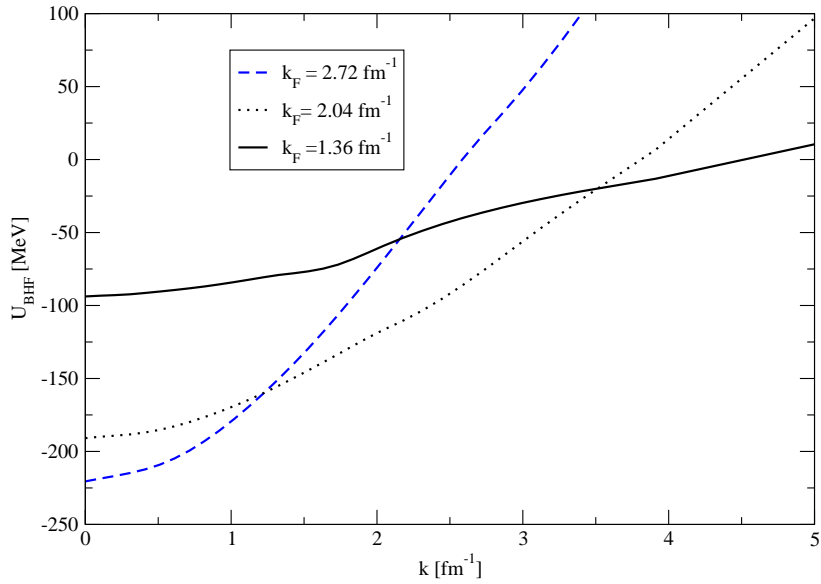}
\end{center}
\caption{The single-particle self-consistent potentials  $U_{BHF}$  (see
({\ref{pot}})) as a function of momentum. Results are
displayed at $k_{F}=1.36$ fm$^{-1}$, $k_{F}=2.04$ fm$^{-1}$ 
and $k_{F}=2.72$ fm$^{-1}$ using the CD-Bonn potential. \label{4}}
\end{figure}

The self-consistent single-particle potential  can be written as follows
\begin{equation}
\label{pot}
U_{BHF}(k)= Re \sum^{BHF}(k,\omega=\varepsilon_{k})\,.
\end{equation}
Fig. \ref{4} displays the single-particle self-consistent potentials as a
function of momentum at different densities using the CD-Bonn potential.
The single-particle potential at $k_{F}$=1.36 fm $^{-1}$ shows a significant
deviation from a parabolic shape in particular at momenta slightly above the
Fermi momentum, see also \cite{frga,gad2,baldo}. It is obvious that such a deviation tends to
provide more attractive matrix elements of $G$ in evaluating the self-energy
for hole states according to ({\ref{eq:selfbhf}}), which leads to more
binding energy. At high Fermi momentum ($k_{F}= 2.04$ fm$^{-1}$ and
$k_{F}= 2.72$ fm$^{-1}$) this deviation is disappear. 

\section{Self-consistent Green functions}
In the self-consistent Green's function (SCGF) formalisms, the self-energy is
the key quantity to determine the one-body Green's function.
One can calculate the contribution of the
hole-hole terms to the self-energy in a kind of perturbative way \cite{grange}
\begin{equation} 
\Delta \Sigma^{2h1p} (k,\omega) = \int_{k_F}^\infty d^3p \int_0^{k_F}
d^3h_1\,d^3h_2\, \frac{<k,p\vert G\vert h_1,h_2>^2}{\omega +
\tilde\varepsilon_p - \tilde\varepsilon_{h_1} -
\tilde\varepsilon_{h_2}-i\eta}\,.\label{eq:2h1p}
\end{equation}

We assume a single-particle spectrum $\tilde\varepsilon_k$ which
is identical to the self-consistent BHF spectrum, but shifted
by a constant $C_1$, which ensures the self-consistency for $k=k_F$
\begin{eqnarray}
\tilde\varepsilon_{k_F} & = & \varepsilon_{k_F}^{BHF} + C_1 \nonumber\\
& = & \frac{k_F^2}{2m} +Re \hspace{1mm} [  \Sigma^{BHF} (k_F,\omega = \tilde\varepsilon_{k_F}) + 
\Delta \Sigma^{2h1p} (k_F,\omega= \tilde\varepsilon_{k_F})]\,.\label{eq:ebhf1}   
\end{eqnarray}
This shifted single-particle spectrum is also used in the Bethe-Goldstone
equation.

\begin{figure}
\begin{center}
  \epsfig{file=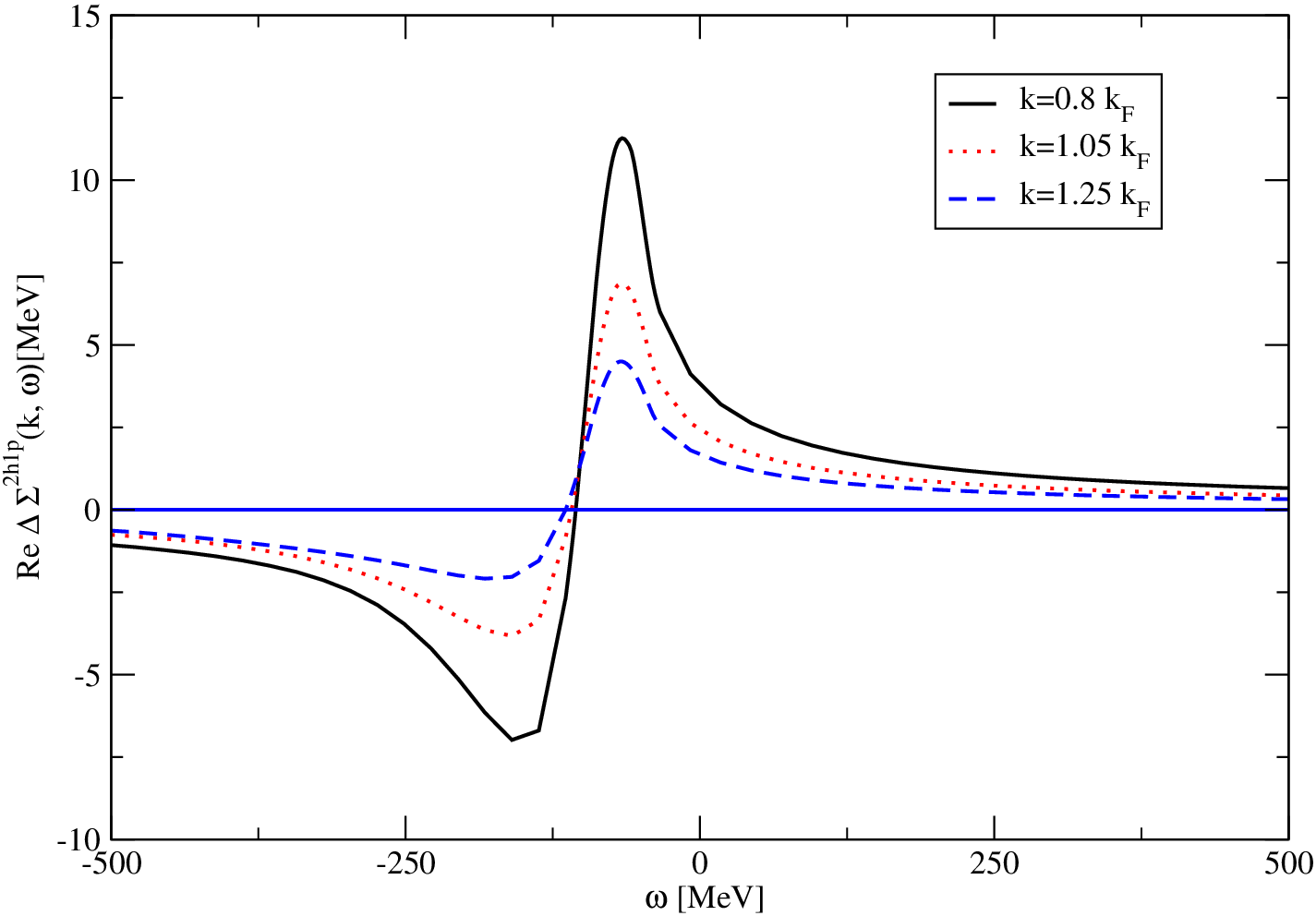,scale=0.45,angle=0}
\end{center}
\caption{The real part of the 2h1p contribution to the self-consistent self-energy as a
  function of $\omega$ (see (\protect{\ref{eq:2h1p}})) 
evaluated for the CD-Bonn potential assuming $k_F$ = 1.36 fm$^{-1}$  
for various momenta $k$. \label{5}}
\end{figure}

\begin{figure}
\begin{center}
  \epsfig{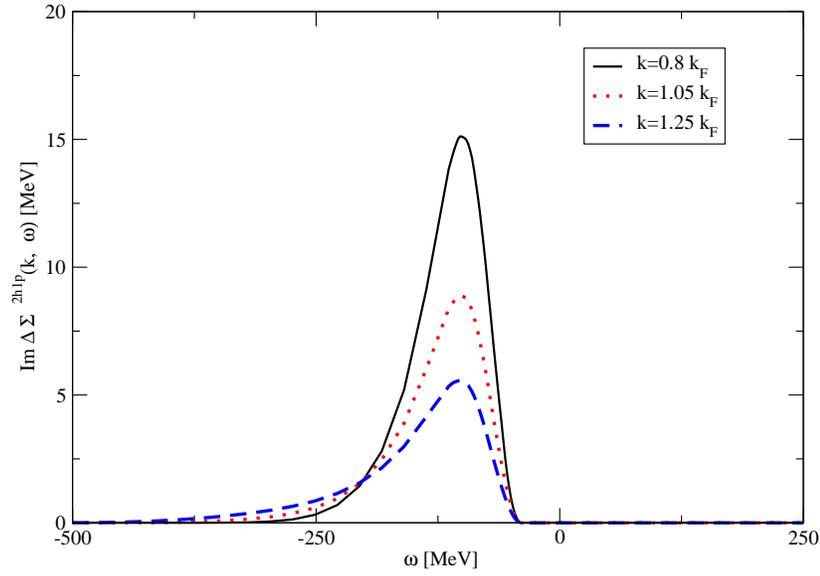}
\end{center}
\caption{The imaginary part of the 2h1p contribution to the self-consistent
 self-energy as a function of $\omega$ (see (\protect{\ref{eq:2h1p}})) 
evaluated for the CD-Bonn potential assuming $k_F$ = 1.36 fm$^{-1}$
for various momenta $k$. \label{6}}
\end{figure}

Results for the two-hole one-particle contribution to the self-consistent self-energy, $\Delta 
\Sigma^{2h1p}$, are displayed in Figs. \ref{5} and \ref{6}, for various momenta
$k$. Fig. \ref{6} illustrates the self-consistent solutions of the imaginary
part of the self-energy, including only 2h1p. 
The imaginary part of $\Delta \Sigma^{2h1p}$ is different from zero only
for energies $\omega$ below the Fermi energy.
This implies that $Im \hspace{1mm} \Sigma^{2h1p}(k, \omega)=0,$ for
$\omega=\varepsilon_{F}$. From (\ref{eq:disper2}) one also observes, noting that
the imaginary part of $\Delta \Sigma^{2h1p}(k, \omega)$ is positive, that for
$k=k_{F}$ the on-shell value of the real part $(\omega=\varepsilon_{F})$ must
be positive. It turns out that this holds for all single particle momenta
below $k_{F}$.
The conservation of the
total momentum in the two-nucleon of the G-matrix in (\ref{eq:2h1p}), $\vec h_1
+ \vec h_2 = \vec k + \vec p$, leads to a minimal value of $\omega$ at which
this imaginary part is different from zero. Due to these limitations the
imaginary part integrated over all energies is much smaller for $\Delta
\Sigma^{2h1p}$ than for $\Sigma^{BHF}$. The real
part of $\Delta \Sigma^{2h1p}$ is related to the imaginary part by a dispersion
relation similar to the one of (\ref{eq:disper2}), connecting the imaginary part
of $\Sigma^{BHF}$ with the particle-particle ladder contributions to the real
part of $\Sigma^{BHF}$. Since the imaginary part of $\Delta \Sigma^{2h1p}$ is
significantly smaller than the one of $\Sigma^{BHF}$, the same is true also for
the corresponding real part. The contribution of $\Delta \Sigma^{2h1p}(k,
\omega)$ term is considerably smaller than the contribution of
$\Sigma^{BHF}(k, \omega)$. At first sight this seems to confirm the notation that the effects
of hh-propagation are considerably smaller than those corresponding to
pp-propagation present in G-matrices.

The Lehmann representation of the Green's function for infinite nuclear matter
is given by \cite{lehman}
\begin{equation}
g(k,\omega)=\int^{\varepsilon_{F}}_{-\infty}d\omega^{\prime}\frac{S_{h}(k,\omega^{\prime})}
{\omega-\omega^{\prime}-i\eta}+\int^{\infty}_{\varepsilon_{F}}d\omega^{\prime}\frac{S_{p}
(k,\omega^{\prime})}{\omega-\omega^{\prime}+i\eta} \mbox{,}
\end{equation}
where $S_{h(p)}(k,\omega)$ is the hole (particle) spectral function and we
will discuss it in detail in the next section.

The formal solution of Dyson's equation is particularly simple in an infinite system,
\begin{equation}
g(k,\omega)=\frac{1}{\omega - \frac{k^2}{2m} - \sum(k,\omega)}\, \mbox{.}
\label{eq:green1}
\end{equation}

In this approximation the solution of the Dyson equation leads only to a shift
in the single particle energy according to (\ref{eq:bhf1}). The Green's
function expression for the binding energy is given by \cite{galitski,frga}
\begin{equation}
\frac{E}{A} = \frac{\int d^3k \int^{\varepsilon_{F}}_{-\infty} d\omega S_{h}(k,
    \omega)\frac{1}{2}(\frac{k^2}{2m} +
\omega)}{\int d^3k \hspace{2mm} n(k)} \mbox{.}\label{eq:ebhf}
\end{equation}

Results for the binding energy per nucleon are displayed in Table \ref{1}. 
In this table we have compared results
for the binding energy of nuclear matter using the CD-Bonn potential from different prescriptions; (i)
Hartree-Fock, (ii) BHF (using continuous choice), and 
 (iii) BHF+2h1p (using conventional choice where the
self-energy is calculated self consistently). The continuous choice leads to
an enhancement of correlation effects in the medium and tends to predict
larger binding energies for nuclear matter than the conventional choice.
The conventional choice has been to
ignore self-energy contribution for the particle states completely and
approximate the energies by the kinetic energy only, see (\ref{eq:conv}). In
nuclear matter one observes, that the inclusion of the 2h1p terms in the
self-energy leads to less attractive quasiparticle energies, but more binding
energy. The gain in binding energy due to the 2h1p components in the
self-energy is about 0.5 MeV per nucleon in nuclear matter at saturation
density \cite{frga}. We found that, no large difference in binding energy
between both results, i.e. BHF and BHF+2h1p.

\section{Self-consistent spectral function}

    The physical meaning of the spectral functions is rather simple. The hole
    spectral function $S_{h}(k,\omega)$ is the probability of removing a
    particle with momentum $k$ from the target system of $A$ particles leaving
    the resulting $(A-1)$ system with an energy $E^{A-1}=E_{0}^{A}-\omega$, where
    $E^{A}_{0}$ is the ground state energy of the target. Analogously, the
    particle spectral function $S_{P}(k,\omega)$ is the probability of adding
    a particle with momentum $k$ and leaving the resulting $(A+1)$ system with
    an excitation energy $\omega$ measured with respect to the ground state of
    the $A$ system, i.e., $\omega=E^{A + 1}- E^{A}_{0}$.
 The spectral functions are obtained from the Dyson equation \cite{mahaux}. For the hole
    spectral function this yields the relation to the self-energy given by
\begin{eqnarray}
\label{eq:specth}
S_{h}(k,\omega) &=& \frac{1}{\pi} Im \hspace{2mm} g(k,\omega)
\nonumber \\
&=&\frac {1}{\pi} \frac {Im \hspace{1mm} \Sigma(k,\omega)}
{(\omega -k^2/2m - Re \Sigma(k,\omega))^2 + (Im \hspace{1mm} \Sigma (k,
\omega))^2} \mbox{.}
\end{eqnarray}
The importance of the particle spectral function is to exhibit where the
missing single-particle strength is located in energy. It is obtained by
solving the Dyson equation and is related to the self-energy by 
\begin{equation}
\label{eq:spectp}
S_{p}(k,\omega)= - \frac {1}{\pi} \frac {Im \Sigma(k,\omega)}
{(\omega -k^2/2m - Re \Sigma(k,\omega))^2 + (Im \Sigma (k,
\omega))^2} \mbox{.}
\end{equation}

\begin{figure}
\begin{center}
  \epsfig{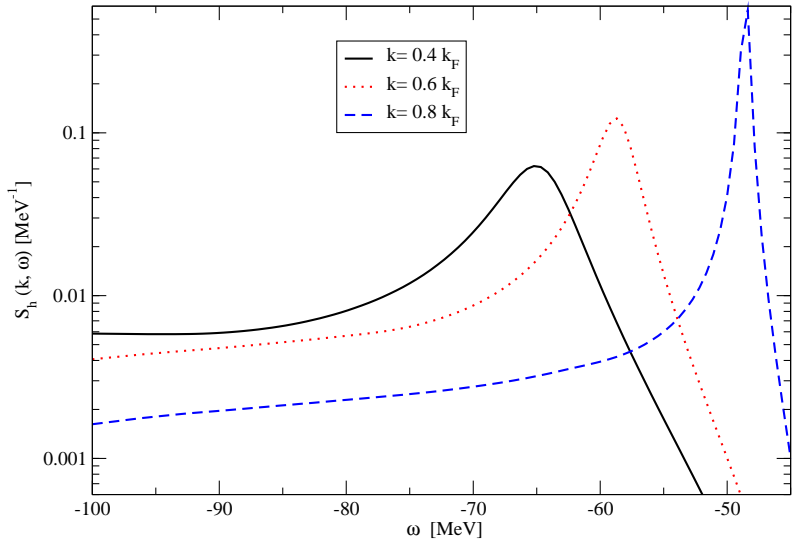}
\end{center}
\caption{  Hole self-consistent spectral functions 
$S_h(k,\omega)$ as a function of $\omega$ for three different momenta in nuclear matter at $k_{F}= 1.36$
fm$^{-1}$ assuming the CD-Bonn potential. 
Note the narrowing of the peak in the spectral function when $k$ gets closer to $k_{F}$.
\label{7a}}
\end{figure}

\begin{figure}
\begin{center}
  \epsfig{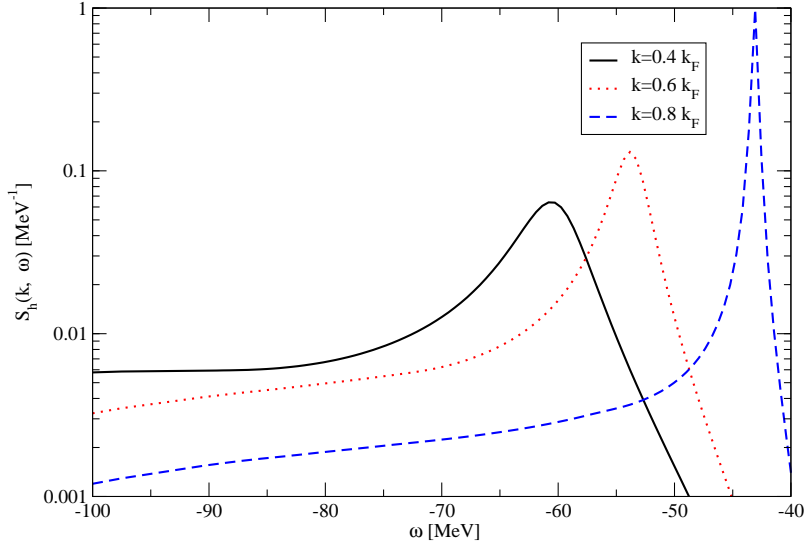}
\end{center}
\caption{  Hole self-consistent spectral functions 
for three different momenta below
$k_{F}$ in nuclear matter at $k_{F}= 1.36$
fm$^{-1}$ using the Bonn C potential.\label{7b}}
\end{figure}

In Figs. \ref{7a} and \ref{7b} a typical example of a hole 
spectral functions, for three different momenta in nuclear matter
at $k_{F}= 1.36$ fm$^{-1}$ as a function of energy. The self-consistent
spectral function (SCSF) calculation is
performed for the CD-Bonn and Bonn C potentials.
The self-consistent particle
spectral function displays a peak at $\varepsilon_{qp}$ because of the
vanishing term in the denominator of (\ref{eq:specth}). Notice that as $k
\rightarrow k_{F}$ this peak becomes extremely sharp due to the vanishing
$Im \hspace{1mm} \Sigma (k, \omega)$ in (\ref{eq:specth}). 

\begin{figure}
\begin{center}
  \epsfig{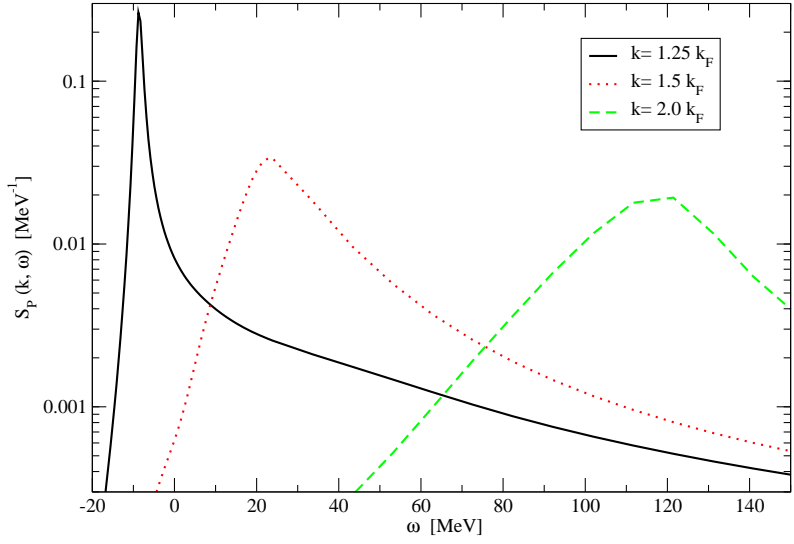}
\end{center}
\caption{ Particle self-consistent spectral functions 
$S_p(k,\omega)$ as a function of energy for various momenta showing a
broadening quasi particle peak with increasing momentum.
The data have been obtained for nuclear matter using the 
CD-Bonn potential with a Fermi momentum $k_F$ = 1.36
fm$^{-1}$ and using gab choice.\label{8a}}
\end{figure}

\begin{figure}
\begin{center}
  \epsfig{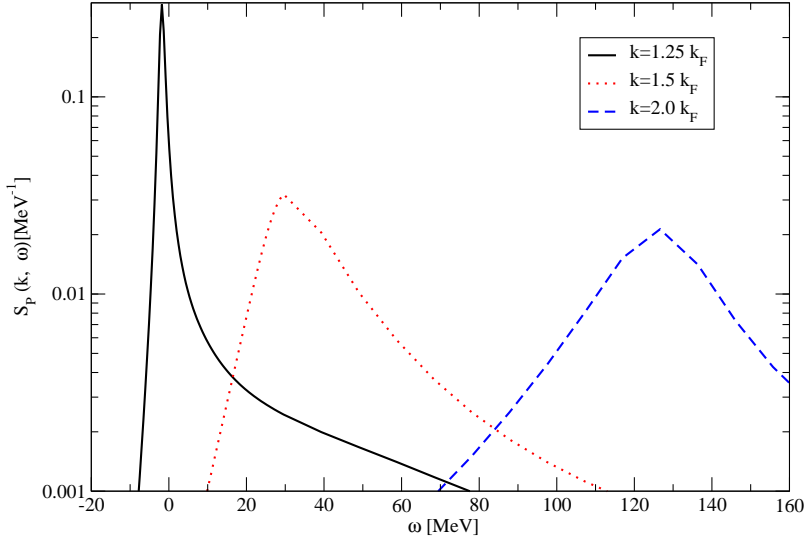}
\end{center}
\caption{ Particle self-consistent spectral functions 
for $k=1.25, 1.5,$ and $2.0 \hspace{1mm} k_{F}$.
The results have been obtained for nuclear matter with a Fermi momentum $k_F$ = 1.36
fm$^{-1}$ and using gab choice, assuming the Bonn C potential. 
\label{8b}}
\end{figure}

In Figs. \ref{8a} and \ref{8b} a typical example of  
a particle spectral functions, for three different momenta in nuclear matter
at $k_{F}= 1.36$ fm$^{-1}$ as a function of energy. The SCSF calculation is
performed for the CD-Bonn and Bonn C potentials.
For momenta larger than $k_{F}$ a quasi-particle peak, which, as usual,
broadens with increasing momentum, can be observed on top of the same high
energy tail. The results therefore display a common, essentially momentum
independent, high-energy tail. The location of single particle strength at
high energy simply means that the interaction has sufficiently large matrix elements.

The quasi-particle contribution can be isolated from the exact single particle
propagator for $k$ close to $k_{F}$ by expanding the self-energy around the
Fermi energy and employing the fact that the imaginary part of the self-energy
behaves like $(\omega-\varepsilon_{F})^{2}$ \cite{luttinger}. The quasi-particle energy is given by
 \begin{equation}
 \varepsilon_{SCGF}^{qp}(k)=\frac{k^{2}}{2m}+Re
 \Sigma(k,\varepsilon^{qp}_{SCGF}(k))\mbox{,} \label{eq:quasienergy}
 \end{equation}
which coincides with (\ref{eq:bhf1}). With these ingredients one obtains
in the limit $k \longrightarrow k_{F}$ a $\delta$-function contribution to the
single particle propagator. Since the imaginary part is typically small around
$\varepsilon_{F}$, this quasi-particle behavior is also obtained for other
momenta close to $k_{F}$.

\begin{figure}
\begin{center}
  \epsfig{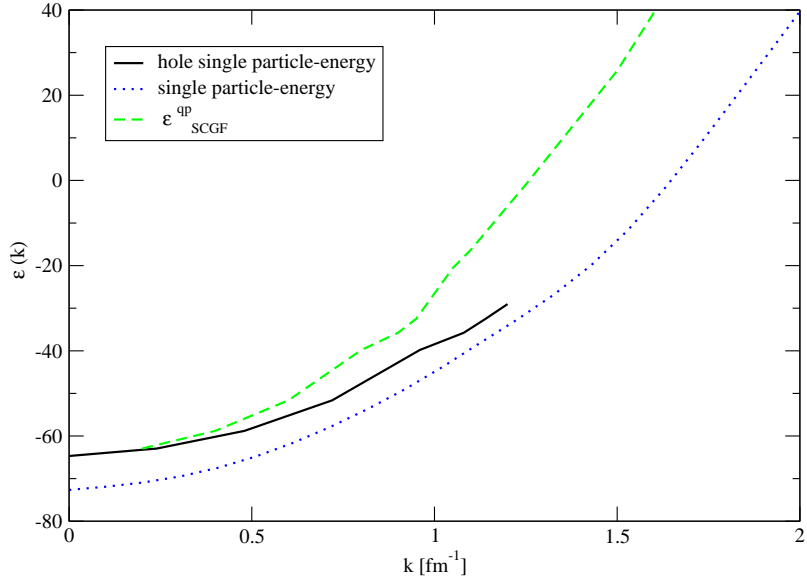}
\end{center}
\caption{The single-particle energy $\varepsilon(k)$ (see ({\ref{eq:conv}}) and
  ({\ref{eq:cont}})) as a function of the momentum at $k_{F}$=1.2 fm$^{-1}$
  using the CD-Bonn potential. For the
  single particle energy for BHF using continuous choice, single particle
  energy below $k_{F}$ for SCGF and
  quasi particle energy ({\ref{eq:quasienergy}}). \label{9}}
\end{figure}

\begin{figure}
\begin{center}
  \epsfig{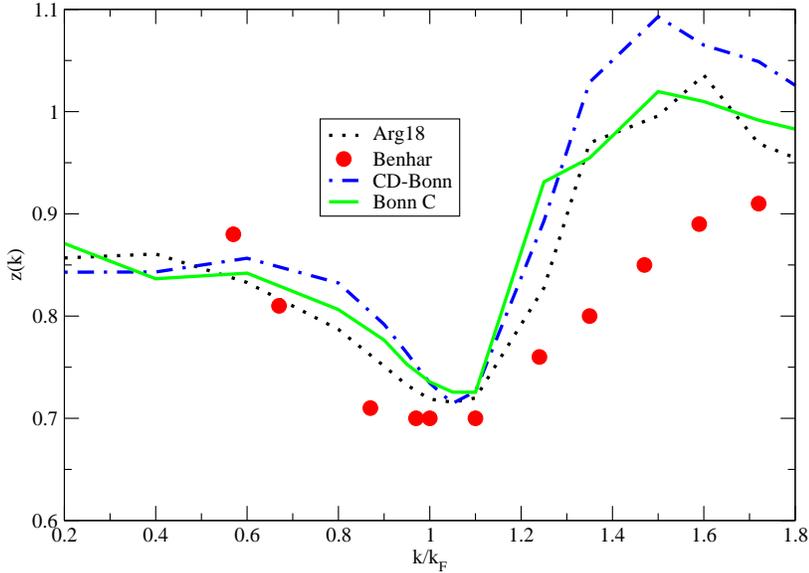}
\end{center}
\caption{Quasi particle strength in nuclear matter at $k_F$=1.36 fm$^{-1}$ (see (\protect{\ref{eq:quasip}}))
  using various models for the NN interaction. For comparison, results
  from the full many-body calculation of Benhar 
  {\it et al.} \cite{ben} (dots) at $k_{F}$= 1.33 fm$^{-1}$ are
  also shown. \label{10}}
\end{figure}

Fig. \ref{9}  displays the single-particle energy 
 $\varepsilon (k)$ (see ({\ref{eq:conv}}) and ({{\ref{eq:cont}}})) as a function 
of the momentum, at $k_{F}$=1.2 fm$^{-1}$ using CD-Bonn potential. In this
 figure we displayed the single particle energies for continuous
  choice, single particle energies below $k_{F}$ for SCGF using the conventional
  choice and the quasi particle energy (\ref{eq:quasienergy}). 
To obtain correct sum rules for momenta close to the Fermi momentum $k_{F}$,
the quasi-particle approximation is used to isolate this part of the
propagator explicity.
The quasi-particle strength $z(k)$ is given as \cite{jeukenne}
\begin{equation}
z(k)=\big[1-\big(\frac{\partial Re \Sigma(k,\omega)}{\partial
  \omega}\big)_{\omega=\varepsilon_{qp}(k)}\big]^{-1} \,,\label{eq:quasip}
\end{equation}
where the quasi-particle strength in nuclear matter at $k_F=1.36$ fm$^{-1}$
  using various models for the NN interaction are shown in Fig. \ref{10}. For comparison, results
  from the full many-body calculation (at $k_{F}= 1.33$ fm $^{-1}$) of Benhar
  {\it{et al}}. \cite{ben} (dots) are also shown in Fig. \ref{10}. 

\begin{figure}
\begin{center} 
  \epsfig{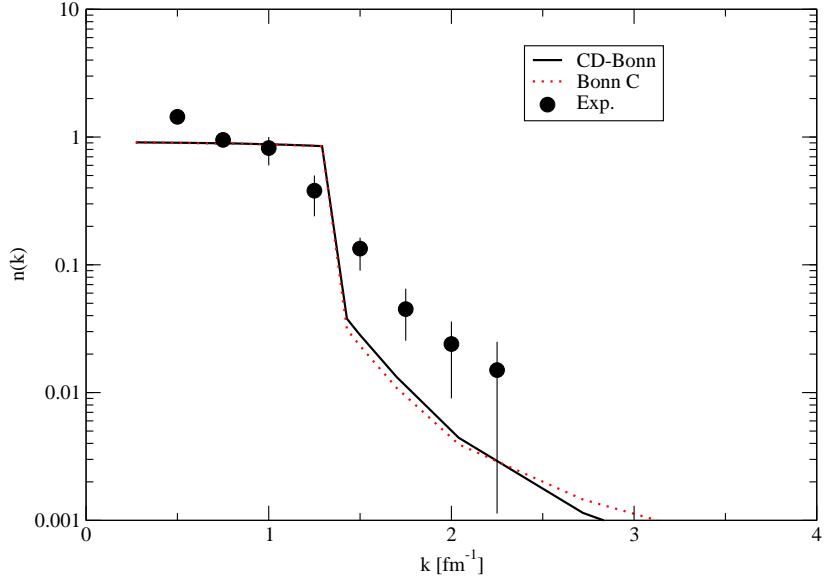}
\end{center}
\caption{Momentum distribution in nuclear matter  calculated from the spectral
  function. Results are given for the CD-Bonn and Bonn C potential at $k_{F}=
  1.36$ fm$^{-1}$. 
The experimental data are from \cite{ciof}. \label{11}}
\end{figure}

Integrating the hole spectral function $S_{h}(k,\omega)$ over all the
   accessible excited states of the (A-1) system, one obtains the occupation
   probability of the single particle momentum $k$ in the correlated ground state.
   The single nucleon momentum distribution in nuclear matter is defined as\\
\begin{equation}
n(k)=\int^{\varepsilon_{F}}_{-\infty}S_{h}(k,\omega)d\omega=
1-\int^{\varepsilon_{F}}_{\infty}S_{P}(k,\omega)d\omega \mbox{.}
\end{equation}
The momentum distribution is calculated self-consistent.
Results are displayed in Fig. \ref{11} for the CD-Bonn and Bonn C potentials at
$k_{F}= 1.36$ fm$^{-1}$. The experimental data are taken from \cite{ciof}. 
The self-consistent occupation probabilities for nuclear matter are in good agreement with 
experimental data.

\section{Conclusion}

In this paper the nucleon properties in the nuclear medium have been
studied using a self-consistent self-energy.
The effects of hole-hole contributions and a self-consistent
treatment within the framework of the Green function approach are
investigated. It is observed that the effects of hole-hole contributions in
binding energies are
not large, this represent a success for the BHF theory.

M\"uther and his collaborators \cite{frga} and Baldo and Fiasconaro \cite{baldo} shown, that the parabolic
approximation for the single particle potential $U(k)$ in the self-consistent
Brueckner scheme introduces an uncertainty of 1-2 MeV near the saturation
density, and therefore it cannot be used in accurate calculations. The full
momentum dependence has to be retained, which prevents the use of a constant
effective mass approximation. In this paper, we have explained that at high densities
the parabolic approximation is valid.

The spectral functions are calculated from the momentum and energy dependent
self-consistent self-energy by solving the Dyson equation.
We found that the nuclear matter self-consistent occupation probabilities are in good agreement with 
the experimental measurements. 

\begin{ack}
One of us (Kh.\ Gad) would like to thank Professor H.\ M\"uther for
useful discussions, guidance and using his computer code. 
\end{ack}

\newpage

\begin{table}
\caption{\label{tab1:}Energy per nucleon for nuclear matter considering three
different densities,  $k_F$ = 1.20, 1.36 and 1.60 fm$^{-1}$, assuming the
CD-Bonn potential. Results are
displayed for HF, BHF with a continuous choice and using the exact Pauli operator \cite{frga}, and
the BHF+2h1p using conventional choice where the self-energy is
calculated self-consistently. All energies are given in MeV per nucleon.} 
\begin{center}
\begin{tabular}{cc|cccc} 
\\
\hline
& $k_F$ [fm $^{-1}$] &
 \multicolumn{1}{c}{HF} & {BHF, exact \cite{frga}} & \multicolumn{1}{c} (\hspace*{-0.2cm}BHF+2h1p)
 \\ \hline
& 1.20 & 2.88 & -15.39 &  -15.34 \\
CD-Bonn & 1.36 & 4.61 & -18.83 &  -18.99 \\
& 1.60 & 10.31 & -22.86 &  -22.78 \\
\hline
\\
\end{tabular}
\end{center}
\end{table}

\end{document}